# Glassy anomalies in the low-temperature thermal properties of a minimally disordered crystalline solid

J. F. Gebbia[1], M. A. Ramos[2], D. Szewczyk[3], A. Jezowski[3], A. I. Krivchikov[4], Y. V. Horbatenko[4], T. Guidi[5], F. J. Bermejo[6], J. Ll. Tamarit[1,*]

[1] Grup de Caracterizació de Materials, Departament de Fisica, EEBE and Barcelona Research Center in Multiscale Science and Engineering, Universitat Politècnica de Catalunya, Eduard Maristany, 10-14, 08019 Barcelona, Catalonia, Spain

[2] Laboratorio de Bajas Temperaturas, Departamento de Física de la Materia Condensada, Condensed Matter Physics Center (IFIMAC) and Instituto Nicolás Cabrera, Universidad Autónoma de Madrid, Francisco Tomás y Valiente 7, 28049 Madrid, Spain

[3] Institute for Low Temperature and Structure Research, Polish Academy of Sciences Okólna 2, 50-422 Wrocław, Poland

[4] B. Verkin Institute for Low Temperature Physics and Engineering of NAS Ukraine, 47 Science Avenue, Kharkov 61103, Ukraine

[5] ISIS Facility, Rutherford Appleton Laboratory, Chilton, Didcot, Oxfordshire OX11 0QX, UK

[6] Instituto de Estructura de la Materia, Consejo Superior de Investigaciones Científicas, CSIC, Serrano 123, 28006 Madrid, Spain

**Abstract**

The low-temperature thermal and transport properties of an unusual kind of crystal exhibiting minimal molecular positional and tilting disorder have been measured. The material, namely, low-dimensional, *highly anisotropic* pentachloronitrobenzene (PCNB) has a layered structure of rhombohedral parallel planes in which the molecules execute large-amplitude in-plane as well as concurrent out-of-plane librational motions. Our study reveals that low-temperature glassy anomalies can be found in a system with minimal disorder due to the freezing of (mostly in-plane) reorientational jumps of molecules between equivalent crystallographic positions with partial site occupation. Our findings will pave the way to a deeper understanding of the origin of above-mentioned universal glassy properties at low temperature.

The quest to clarify the origin of the characteristic features exhibited by the thermal properties of glassy matter has concentrated significant research efforts since such thermal and transport anomalies were first reported some 45 years ago [1,2]. The basic experimental facts are now understood on purely phenomenological grounds as a consequence of the disorder-induced appearance of low-energy modes able to interact with long-wavelength, heat-carrying excitations. Such vibrational features influence the behavior of the low-temperature specific heat $C_p$ of glasses, which has been found to be large and almost linearly dependent on temperature, and the thermal conductivity $\kappa(T)$ which shows a quadratic temperature dependence and displays values some orders of magnitude lower than those found in their crystalline counterparts [1,2]. A particularly puzzling aspect of the properties of disordered matter, including a substantial number of disordered crystals and even quasi-crystals, refers to the quantitative similarity of properties governing the heat transport features, such as the ratio of the phonon wavelength to its mean-free path [3]. The available data shows that, apart from a few notable exceptions, properties such as the thermal conductivity and acoustic attenuation display quantitative similarities independent of the chemical nature or preparation procedures of the disordered samples.

Most properties of amorphous solids at low temperature have been described by a purely phenomenological model known as the Standard Tunneling Model (STM) [3-5]. The model postulates the ubiquitous existence of small groups of entities, arising as a consequence of some intrinsic disorder, which can tunnel between two configurations of very similar energy, namely tunneling two-level systems (TLS) [6]. The picture however seems to be incomplete since it cannot account for all the "universal" behavior mentioned above, nor is able to explain other phenomena observed at millikelvin temperatures [7-10].

Despite efforts carried out during the last decades, not much is known about the microscopic nature of entities responsible for the observed behavior, exception made of a few well characterized systems [11]. Progress in understanding these phenomena may arise from the emergence of new experimental techniques [12,13], as well as from the study of materials where the effects of atomic-scale disorder can be quantified in detail. Here we focus on molecular crystals having some kind of built-in disorder. This may consist of a complete disorder of the molecular orientations while the periodicity of the network is preserved, as in plastic or orientationally-disordered crystals [3,14-20], or some more restricted types of disorder such as occupational disorder exhibited by otherwise fully-ordered crystals [21-28].

Besides the TLS-dominated range below 1 K, other glassy features arise above some 1 K. For instance, glasses also show a broad maximum in $C_p/T^3$ at about 3-10 K and a plateau in $\kappa(T)$ within the same temperature range. In addition, the origin of the characteristic features shown by the vibrational density of states (VDOS), $g(\omega)$, within the region about 1 THz (about 4 meV), where glasses show a clear excess over the behavior as predicted by Debye's continuum theory, remains to be well understood on a microscopic scale. Such an excess of low-frequency states appears as a broad peak in the reduced g(ω)/ω$^2$ at a few meV, usually referred to as the "boson peak". It also corresponds to the maximum found in $C_p/T^{\,3}$ [29], a feature ubiquitously observed in amorphous solids and also in disordered crystals exhibiting glassy behavior [15-20].

Several theories and models [30-36] have been proposed to account for the boson peak and related features in thermal and dynamical properties of glasses. We will not discuss them here, but just mention the phenomenological soft-potential model (SPM) [30,31], which can be considered as an extension of the abovementioned STM and provides analytical expressions to assess experimental data [37]. Specifically, the SPM postulates the coexistence in glasses of acoustic phonon-like lattice vibrations with quasilocalized low-frequency ones. These vibrational modes are sometimes called propagons, diffusons and locons depending on whether they are propagating phonon-like, diffusion or localized modes, respectively [38].

Here we study a minimally disordered crystalline system in which only a few degrees of freedom make it to depart from a fully ordered state. The chosen system is a quasi-planar hexasubstituted benzene derivative, pentachloronitrobenzene (PCNB, $C_6Cl_5NO_2$), that exhibits a layered structure of rhombohedral (space group $R\bar{3}$) parallel planes in which the molecules can rotate around a six-fold-like axis [21,39-42]. Nevertheless, the molecular dipole has two nonzero (in-plane and a small out-of-plane) components with a highly anisotropic dynamical coupling. This gives rise to two reorientational dynamical processes, associated with reorientational motions of the larger dipole component within the (001) planes of the hexagonal structure and with the small dipole component fluctuations around the *c* hexagonal axis [21]. Therefore, PCNB is a *highly anisotropic* solid that can be considered a crystal with a low-dimensional orientational disorder, which attracted a lot of attention in the early times [41,42]. Specifically, data for PCNB crystals did show remarkable departures from expected correlations between the observed moments and the bond moments of the constituent molecules. In addition, spectroscopic measurements [42] revealed that, curiously enough, the root mean square torque acting on a librating molecule decreases with decreasing temperature despite a significant contraction of the crystal lattice. The pioneering study of Weiss highlighted the molecular pseudo-symmetry of the halogenobenzenes as the main cause for the subtle disorder [43] and some further studies accounted for that orientational disorder [44,45]. According to recent works [39,40], the constituent PCNB molecules exhibit remarkably large (temperature-independent) atomic displacements giving rise to a highly distorted crystal structure. Such motions are allowed because of anomalously large values for some of the intermolecular contact distances within a crystal structure which exhibits molecular positional and tilting disorder. The net effect of such distortions is to generate strain fields which are felt by the librating dipoles.

We have taken this material as a model system in our study on glass properties since the absence of a low-temperature fully ordered ground state as well as the presence of significant distortion fields, may lead to the formation of tunneling systems. At this point it is worth recalling the difference between the behavior exhibited by the material of interest and its chemical analogue pentamethylbromobenzene [45], which does show a transition into a fully-ordered crystal lattice. Hence our expectation was to test whether PCNB crystals may constitute a molecular material which could enable the identification of the microscopic entities responsible for glassy behavior as achieved in the well-studied case of alkali-halide crystals [11]. It would thus be pertinent to carry out measurements of the specific heat down to temperatures below 1 K and check whether any clear power-law dependence with temperature, with an exponent close to unity, could be found. In addition, if this were the case, one would also expect the material to exhibit a low thermal conductivity due to the operation of strong phonon-scattering channels and thus the thermal conductivity could become comparable in magnitude to that exhibited by structural glasses.

In Fig. 1 we show our measurements of the specific heat as a function of temperature down to 0.3 K. Within the $C_p/T^3$ representation, data clearly show the boson peak as a broad maximum at about $T_{BP}$ = 4.8 K, with a height of $(C_p/T^3)_{max}$ = 7.85 mJ mol$^{-1}$ K$^{-4}$. In the inset of Fig. 1, the specific heat of PCNB at higher temperatures is depicted. A small but clear jump in $C_p$ is observed at about 190 K which corresponds to the glass-like transition of this disordered crystal [46], where the in-plane reorientational degrees of freedom freeze-in.

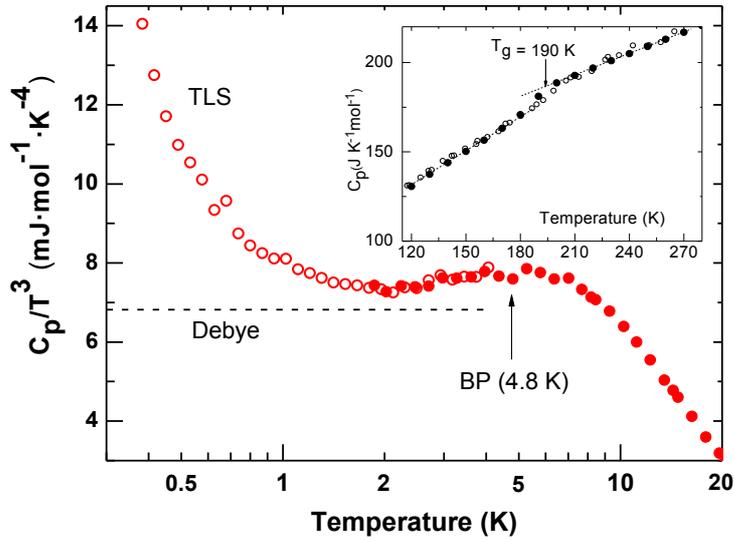

Figure 1. Debye-reduced specific heat data $C_p/T^3$ for the low-temperature crystalline phase of PCNB. Different symbols indicate different experimental runs. The arrow marks the position of a shallow maximum (boson peak) observed at T=4.8 K. The dashed curve indicates the Debye coefficient obtained from a SPM parabolic fit (see Fig. 3). The upturn of the curve below 1 K is a signature of a quasilinear contribution to the specific heat ascribed to the existence of glassy two-level systems. Inset: Specific heat data $C_p(T)$ at higher temperatures (including data, empty symbols, from Tan et al. [46]) showing at around 190 K a small jump in $C_p$ corresponding to the glass-like transition of this disordered crystal.

The specific heat data are fully consistent with our estimate of the spectral frequency distribution, g(ω), measured by inelastic neutron scattering (INS) experiments at the MARI spectrometer at ISIS, shown in Fig. 2. The measurement was limited to frequencies above some 2 meV due to instrument limitations. The reduced VDOS, $g(\omega)/\omega^2$, shown in the inset, is consistent with the presence of a weak feature or "boson peak" centered at about 2 meV superimposed to the Debye VDOS, $g(\omega) \propto \omega^2$. It is worth mentioning that the crystal nature of the material enables an unambiguous identification of features appearing in g(ω) to discrete normal modes and in fact, the observed extrema match the librational frequencies derived from the normal mode analysis of the atomic displacements [40], yielding 4.26 meV, 5.36 meV and 5.98 meV, as well as a peak at 8.55 meV assigned to molecular torsional oscillations.

As mentioned above, the most distinctive fingerprints of the glassy state at low temperature are the presence of a quasilinear contribution in the specific heat, $C_p \propto T$, and a quadratic dependence of the thermal conductivity, $\kappa \propto T^2$ [1, 6]. The present results concerning the specific heat of the disordered crystalline PCNB solid display the same striking behavior as that shown by fully amorphous solids. This is shown in Fig. 3 where the specific heat is plotted as $C_p/T$ vs $T^2$ below about 4 K. Notice that such a linear contribution is also revealed by the clear upturn of the curve below 1 K in Fig. 1.

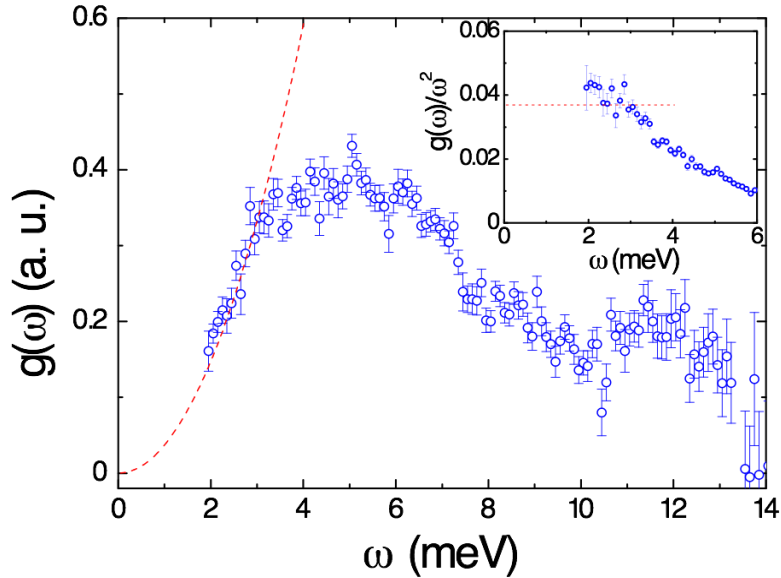

Figure 2. Vibrational density of states g(ω) for the low-temperature crystalline phase of PCNB derived from INS measurements. The dashed red curve corresponds to the Debye approximation valid at the lower frequencies. Inset: Reduced VDOS [$g(\omega)/\omega^2$] for the data in the main figure.

To provide a direct comparison of the strength of the thermal anomalies in PCNB with those shown by amorphous matter, a least-squares linear fit of $C_p/T$ vs $T^2$ data at the lowest temperatures (T < 1.4 K) is performed (see Fig. 3). This provides the coefficients for $C_p = C_{TLS} T + C_D T^3$. We obtain $C_{TLS}$=1.0 mJ mol$^{-1}$ K$^{-2}$, a coefficient customarily ascribed to tunneling two-level states following the STM, and the Debye coefficient $C_D$=7.0 mJ mol$^{-1}$ K$^{-4}$ (the slope of the curve).

For temperatures above 2 K the experimental data deviates from that linear fit, since the latter does not take into account the contribution associated with finite frequency excitations [37]. In addition, for temperatures still below the maximum in $C_p/T^3$ the data can be accounted for by the addition of a further term which is usually identified with excess quasilocalized vibrations (*soft modes*) which follow $g_{sm}(\omega) \propto \omega^4$ and hence $C_{p,sm} \propto T^5$ [31]. A much better description and numerical evaluation of the data can therefore be performed [37] by considering a simple parabolic fit of $C_p/T$ vs $T^2$ data, associated with the whole range of temperatures, that is given by:

$$C_p = C_{TLS} T + C_D T^3 + C_{sm} T^5 \quad (1)$$

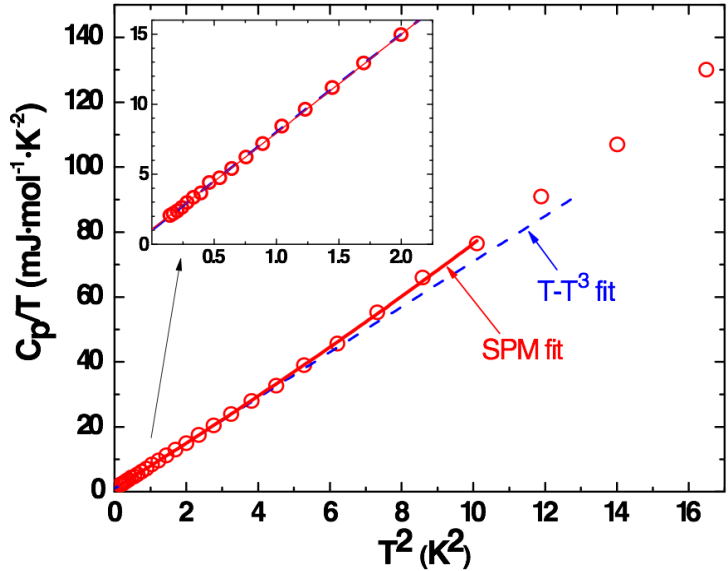

Figure 3. Specific heat data plotted as $C_p/T$ vs $T^2$ at low temperatures for crystalline PCNB. The dashed (blue) curve shows a simple least-squares linear fit to clearly determine the linear coefficient in the zero-temperature limit, whereas the solid (red) line shows the result from a SPM parabolic fit in a wider temperature range (see text). Inset: Amplification of the data and fitting curves at the lowest temperatures.

To be consistent, the temperature range chosen to perform this quadratic fit should be between zero temperature and roughly the midpoint between the minimum and the maximum of the $C_p/T^3$ curve [18]. By doing so, the SPM fit (shown by a solid line in Fig. 3) gives: $C_{TLS}$ = 1.06±0.11 mJ mol$^{-1}$ K$^{-2}$, $C_D$ = 6.82±0.08 mJ mol$^{-1}$ K$^{-4}$, and $C_{sm}$ = 0.072±0.009 mJ mol$^{-1}$K$^{-6}$. These coefficients are more reliable than those from the simpler linear fit at the lowest temperatures shown before, but the reasonable agreement in both $C_{TLS}$ and $C_D$ coefficients supports the consistency of our data analysis. Here it is worth emphasizing the rather small value found for the contribution of soft modes, $C_{sm}$. The strength of this term is in turn related to a frequency contribution to $g_{sm}(\omega) \propto \omega^4$. Its rather small value is thus consistent with the weak feature ascribable to glassy excess modes (boson peak) appearing at the VDOS shown in Fig. 2 for frequencies about some 2 meV. Nonetheless, in glasses with much stronger boson peaks the former linear fit has been shown [37] to provide wrong TLS and Debye coefficients, in contrast to the latter parabolic SPM fit, which we will employ hereafter.

It is also worth mentioning here that both the heat capacity jump and the low intensity of the boson peak are noticeably smaller than those found in other (typically isotropic) orientationally-disordered crystals such as ethanol, cyclohexanol, Freon 112 or Freon 113, the most fragile orientational glass reported so far [15–20]. The molar density of TLS in our PCNB crystal ($C_{TLS}$ = 1.06±0.11 mJ mol$^{-1}$ K$^{-2}$) is however very similar to those previously found in molecular glasses of ethanol (normal and deuterated) or propanol (*n*− and iso−) [47], as well as of different isomers of butanol (*n*−, sec−, and iso−) [48]. Interestingly, also orientationally-disordered crystals of ethanol exhibited $C_{TLS}$ ~ 1 mJ mol$^{-1}$ K$^{-2}$ [47]. In addition, from the Debye coefficient (indicated by a dashed horizontal line in Fig. 1), a *molecular* [48] Debye temperature $\Theta_D$=66 K is obtained, slightly lower than those found in typical molecular glasses, despite formally being a crystal.

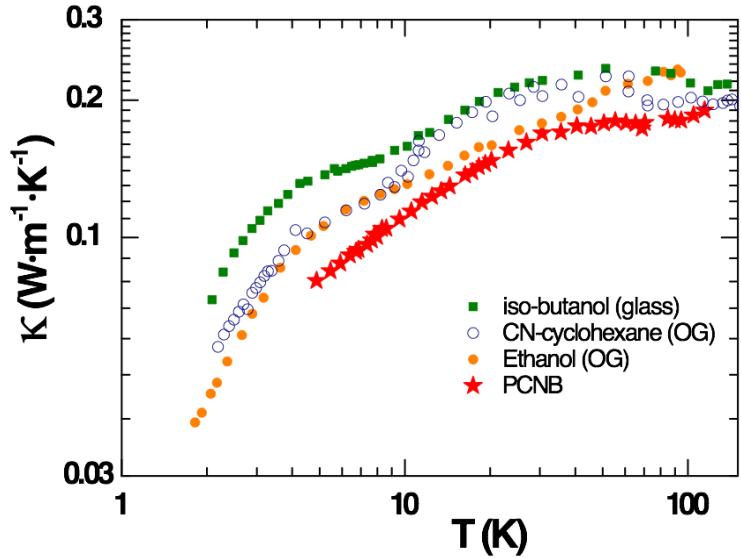

Figure 4. Log-log plot of the thermal conductivity as a function of temperature for PCNB low-dimensional glassy crystal (red stars). For comparison, the orientational glasses of ethanol (orange solid circles) [49] and CN-cyclohexane (open circles) [50], and the canonical glass of iso-butanol (green squares) [48], are also shown.

Our thermal conductivity data as measured over a temperature range from 4 K upwards are shown in Fig. 4 together with data pertaining two orientational glasses and a canonical glass (iso-butanol), using a double logarithmic scale. The similarity of the thermal conductivity at about 10 K for all these glassy materials, without any scaling, is indeed remarkable. The lack of data below 4 K hinders a further, fully detailed analysis.

For temperatures higher than 4 K up to ~ 40 K, thermal transport in disordered matter is ascribed to the action of heat carriers transporting heat via diffusive random-walk steps or "diffusons". Within that range, the classical theory of thermal conductivity is able to explain the experimental observations associated to an increase of the specific heat while both the speed of sound and the phonon mean-free path remain temperature independent. Heat transport processes become strongly scattered for temperatures between 40 K and 100 K thus leading to the observable plateau displayed by our data in Fig. 4. In this temperature range, the lowest frequency librational modes sampled by our measurement of g(ω) described above, will become thermally populated and therefore will become able to strongly scatter heat-carrying excitations. The motions executed by these low-frequency modes are known to be temperature-independent and therefore may well constitute a microscopic realization of the "soft modes" concept.

Our results regarding a stable, highly anisotropic crystalline material can be contextualized with recent discussions about the absence of TLS excitations in some materials that are considered notable exceptions to the universality of glassy behavior [51-53]. The results recently found in one of those so-called ultrastable glasses [53] suggested that the anisotropic and layered structure of this glassy material, induced by the conditions used to prepare those vapor-deposited glasses, would be the main source of the deviation from the universal behavior. Nonetheless, the particular anisotropic growth of the studied ultrastable indomethacin glass also appears to imply a decrease of free hydrogen bonds and an enhancement of π−π interactions between chlorophenyl rings. Indeed, loss of stability and concomitant increase of water absorption seemed to facilitate recovery of the three-dimensional dynamic network and hence of the presence of TLS [53]. Our current findings suggest that such reported anomalies are best explained as consequences of those specific structural peculiarities of the material, and that anisotropy itself does not necessarily prevent the existence of TLS in glassy systems. Further experiments in other possible anisotropic glass-forming materials would be most interesting to shed light on this intriguing issue.

Finally, the relevance of our results can also be assessed by consideration of the thermal conductivity data displayed in Fig. 4 for a set of different molecular materials. Data analyzed so far [54,55] show that the strength of the *C* parameter governing the strength of phonon scattering by low-energy excitations is of the order of $10^{-4}$ for all the studied glasses and the characteristic energies *W* become of the order of a few kelvins. Available data for PCNB indicates that its relevant parameters show values comparable to those of previously analyzed materials. Therefore we see that the effect of glasslike excitations on this transport property becomes "saturated" in the sense that the absolute values of these quantities become independent of the amount and the kind of disorder [56]. At this point it is worth considering the current results within the broader context of localized excitations within defective crystals. In fact, the presence of two-level systems in some molecular crystals at low temperatures was known [57] well before a plausible model able to account for the universal glassy anomalies was formulated [4,5]. In more recent times, the issue has regained attention due to results from experiments by single-molecule spectroscopy [58]. The phenomena there sampled however seem to be of a different nature from those exhibited by canonical glasses. In fact, recent reports have questioned the ability of the technique to provide a picture consistent with the wealth of data already available on glassy dynamics. Moreover, more detailed results on partially- or fully-amorphous hosts [59,60] have shown that the experimental results could not be sensibly analyzed in terms of the Standard Tunneling Model. These microscopic experiments, as well as abovementioned macroscopic (ensemble-averaged) ones [51-53], cast doubt on the truly universality of glassy behavior and its interpretation by the STM. Studying, both experimentally and theoretically, different systems with distinct degrees of disorder seems essential to understand the reach of such universality, and the reason for their exceptions.

To summarize, our current results show that a crystal with a minimal amount of disorder displays a linear contribution to the specific heat at very low temperatures, as well as a glassy behavior of its thermal conductivity at intermediate temperatures, the absolute values of which are comparable to other materials, either within their orientational-glass or fully amorphous states. Nevertheless, both the magnitude of the specific-heat jump at the glass-like transition and the height of the boson peak are noticeably smaller (but not negligible) than those reported for systems in which only orientational degrees of freedom are frozen-in [15–20].

This work has been supported by the Spanish MINECO through projects FIS2014-54734-P, FIS2014-54498-R and MAT2014-57866-REDT and "María de Maeztu" Program for Units of Excellence in R&D (MDM-2014-0377) and the National Science Centre (Poland) grant No. UMO-2013/08/M/ST3/00934. We also acknowledge the Generalitat de Catalunya under project 2014SGR-581 and the Autonomous Community of Madrid through program NANOFRONTMAG-CM (S2013/MIT-2850).